\documentclass[sigconf,natbib=true]{acmart}
\usepackage{enumitem}
\usepackage{dirtytalk}
\usepackage{subcaption}
\usepackage{textgreek}
\usepackage{multirow}

\usepackage{amssymb}   % for \checkmark
\usepackage{pifont}    % for \ding commands

\newlist{questions}{enumerate}{2}
\setlist[questions,1]{label=\textbf{RQ:},ref=\textbf{RQ}}

\newtheoremstyle{boldhyp} % name
    {0em}                    % Space above
    {0em}                    % Space below
    {\itshape}                    % Body font
    {\parindent}                  % Indent amount
    {\bfseries}           % Theorem head font
    {.}                   % Punctuation after theorem head
    {.3em}                % Space after theorem head
    {}                    % Theorem head spec (can be left empty, meaning ‘normal’)

\theoremstyle{boldhyp}

\AtBeginDocument{%
  }

\copyrightyear{2026}
\acmYear{2026}
\setcopyright{cc}
\setcctype{by}
\acmConference[CHIIR '26]{2026 ACM SIGIR Conference on Human Information Interaction and Retrieval}{March 22--26, 2026}{Seattle, WA, USA}
\acmBooktitle{2026 ACM SIGIR Conference on Human Information Interaction and Retrieval (CHIIR '26), March 22--26, 2026, Seattle, WA, USA}
\acmPrice{}
\acmDOI{10.1145/3786304.3787917}
\acmISBN{979-8-4007-2414-5/2026/03}

\begin{document}

%%
%% HEADER ---------------------------------------------------------

%%
%% TITLE

\title{Cleo: A Transparent and Controllable Chatbot for Conversational Commerce}

% \title{Cleo: Separating Decision Policy from Language Generation in Conversational Product Search}

\author{Kevin Schott}
\email{kevin.schott@gesis.org}
\affiliation{%
 \institution{GESIS – Leibniz Institute for the Social Sciences}
 \city{Cologne}
 \country{Germany}
 }
 
\author{Jan Lattenkamp}
\email{jan.lattenkamp@gesis.org}
\affiliation{%
 \institution{GESIS – Leibniz Institute for the Social Sciences}
 \city{Cologne}
 \country{Germany}
 }

\author{Daniel Hienert}
\email{daniel.hienert@gesis.org}
\affiliation{%
 \institution{GESIS – Leibniz Institute for the Social Sciences}
 \city{Cologne}
 \country{Germany}
 }

\author{Dagmar Kern}
\email{dagmar.kern@gesis.org}
\affiliation{%
 \institution{GESIS – Leibniz Institute for the Social Sciences}
 \city{Cologne}
 \country{Germany}
 }
 
%%
%% SHORT AUTHORS
\renewcommand{\shortauthors}{Schott et al.}

\begin{abstract}
We demonstrate Cleo, a transparent and controllable conversational product advisor that addresses the challenges of opacity, unpredictability of LLMs, and the complexity of comparisons in conversational commerce. With our chatbot system, we make four contributions: First, we introduce transparency by prompting the LLM to reflect on interpreted user needs, while an auditable ranking mechanism reveals loss values per attribute, explaining ranking decisions. Second, we propose controllability through a hybrid architecture separating deterministic ranking from language generation. A ranker applies categorical filters and numeric loss functions over 3,638 product specifications. Meanwhile, a constrained LLM generates grounded descriptions constrained to catalog evidence, thus mitigating the risk of hallucinated or persuasive content. Third, we provide decision support in the form of natural-language comparisons and a highlights feature. These aim to reduce mental workload by contextualizing specifications relative to user needs. Fourth, we contribute an extensible experimental system for IR and HCI researchers, as well as practitioners of conversational search and recommendation. Unlike traditional faceted search or opaque LLM-only recommenders, our approach allows for fluid conversation while maintaining algorithmic transparency. In a live demonstration, attendees will experience information needs elicitation and reflection, conversational refinement with real-time re-ranking, inspection of per-attribute loss explanations, and AI-generated multi-item comparisons. The system aims to advance the design of transparent and controllable conversational systems that provide support for decision-making during online product search.
\end{abstract}

% One-line problem + what your system lets users do (steer via conversation, audit, compare) + what will be shown live. Keep claims modest; promise transparency over accuracy gains. Many CHIIR demos front-load a crisp abstract, then go straight into an architecture or UI overview.

%%
%% ARTICLE TYPE
\acmArticleType{Research}

%% CCSXML



\maketitle

%%
%% MAIN BODY ------------------------------------------------------

%% Framing: A transparent prodcut advisor (exposing the ranking/lossvalues and informing the user how their (vague) needs were interpreted) that supports decision-making; we keep the LLM in check and aim to enhance reliability thorugh safeguards (by sanitizing the LLMs output in the extraction step and hard coding the rules for the highlights and comparisons)

\section{Introduction \& Contribution}
\label{sec:motivation}

% Introduction
Conversational commerce, the buying activity conducted via a digital assistant, aims to provide chat- or voice-based product advice that mimics a human shop assistant: the assistant elicits needs and preferences based on which it recommends suitable items~\cite{Messina,Tuzovic2018,ConversationalCE,Tsagkias2021}. Since these assistants act as interactive, socially present decision-making aids, they can foster trust in e-commerce platforms, which is a prerequisite for purchase intentions~\cite{Virdi2020,Gefen2003}.

% Context
The process of online shopping with a digital assistant combines search and recommendation~\cite{Zhang2018,Radlinski2017,Schott2024}. Contemporary conversational recommenders promise multi-turn preference elicitation beyond one-shot results lists, yet dialogic answers often hide ranking logic and underlying evidence~\cite{Jannach2022,Gao2021}. For conversational search, Radlinski and Craswell~\cite{Radlinski2017} introduced the concept of \textit{revealment}: helping users express or discover their needs (user revealment) and exposing the system's capabilities and corpus (system revealment). We adapt their notion of system revealment to also encompass surfacing internal decision-making, allowing users to assess system capabilities. Łajewska et al.~\cite{Lajewska2024} show that in conversational information systems, where details like source attribution and model confidence are often concealed, users struggle to detect factual errors and biases in responses.

% Utility of CRS/conversational search systems for product search
Many shopping tasks are faceted and multi-attribute, with users often uncertain which attributes matter or how to reference them~\cite{Radlinski2017}. Mixed-initiative conversation supports piecewise elicitation and lightweight teaching~\cite{Radlinski2017}, while critiquing enables attribute-level refinement by the user through natural-language feedback~\cite{Chen2012}. In addition, clarifying questions from the assistant can reduce ambiguity before showing results, improving retrieval effectiveness~\cite{Aliannejadi2019}.

% Problem 1 -- Comparison challenge
However, comparing and evaluating retrieved products remains challenging. Users frequently open multiple browser tabs for side-by-side comparison~\cite{Schott2025}, while existing comparison features typically present specification tables requiring manual interpretation of technical details and trade-offs. Personalized natural-language comparisons provided by conversational assistants offer the potential to reduce mental demand and support informed decision-making.

% Problem 2 -- Black-box recommendations erode trust and scrutability
These benefits are undermined when recommenders remain black boxes, reducing scrutability and potentially affecting trust~\cite{Tintarev2012}. 
%Tintarev and Masthoff~\cite{Tintarev2015,Tintarev2012} identify seven aims of explanations in recommender systems--effectiveness, satisfaction, transparency, scrutability, trust, persuasiveness, efficiency--which are often in tension. User-centered evaluations show that explanations can enhance decision effectiveness~\cite{Tintarev2012,Pu2011}. 
In conversational commerce, disclosing what was understood has been shown to increase perceived competence and engagement~\cite{Papenmeier2023}, while natural-language explanations linking user utterances to product attributes can improve perceived transparency~\cite{Schott2024}. However, explanation quality matters: Łajewska et al.~\cite{Lajewska2024} demonstrate that while high-quality explanations of sources, model confidence, and limitations tend to raise perceived response usefulness and explanation ratings, inaccurate explanations can significantly reduce perceived usefulness. For e-commerce systems, Tsagkias et al.~\cite{Tsagkias2021} point out that explanations can increase transparency by enabling users to understand what data from their input is being processed and how the search or recommendation mechanism works.

% Problem 3 -- LLM outputs can be helpful but are unpredictable; we need steering & controls
Large language models (LLMs) introduce additional complexity. Supervised fine-tuning leaves models prone to undesired behavior, while reinforcement learning from human feedback (RLHF) substantially improves but does not guarantee instruction-following~\cite{Ouyang2022,Wang2024}. LLM-based agents can exhibit overconfidence~\cite{Sun2025,Wen2024} or employ persuasive tactics~\cite{Liu2025,Rogiers2024}. Lin et al.~\cite{Lin2025} advocate hybrid approaches, noting that LLM-only recommenders without domain-specific fine-tuning can lag in performance and introduce latency. While retrieval-augmented generation (RAG) provides grounding~\cite{Grainger2025,Withorn2025}, challenges such as hallucinations, biases, and incorrect instruction-following persist~\cite{Li2024,Zhang2025comprehensive,Chen2024}.

% Contribution to address the identified problems
We present a chatbot, acting as a conversational product advisor, addressing these challenges through four contributions:

\textbf{(1) Controllability:} We demonstrate a hybrid architecture separating deterministic ranking from language generation, mitigating hallucinated or persuasive content.

\textbf{(2) Transparency:} We operationalize system revealment through transparent and verifiable ranking.
%with per-attribute loss exposure and prompting that reflects back interpreted user needs.

\textbf{(3) Decision support:} We provide constrained natural-language comparisons and a highlights feature designed to reduce cognitive load by contextualizing specifications in relation to user needs.

\textbf{(4) Experimental system:} We contribute an open-source extensible system designed to support research on transparency, control, and decision support in conversational search and recommendation\footnote{The current, work-in-progress version of Cleo can be tested via \url{https://multiweb.gesis.org/vacos5/}.}.

This approach provides a practical balance between fully automated LLM-based recommenders and traditional faceted search. It reduces comparison complexity through transparency and improves interpretability with verifiable rankings. It furthermore limits LLM unpredictability through constrained text generation.

\section{User Interaction}
\label{sec:ui}
We implemented and instantiated Cleo as a conversational agent (chatbot) designed to support users in finding a suitable laptop. While the current implementation focuses on laptops, the system can easily be adapted to other product domains characterized by a set of structured attributes. We selected laptops as our use case because they represent a complex yet familiar product category, involving multiple comparable features such as performance, portability, and price.

Cleo supports both text-based and voice-based input, allowing users to express their information need through their preferred modality. The system includes text-to-speech functionality to vocalize the chatbot's responses, and allows users to edit their most recent message. Through natural conversational interaction, users engage in a mixed-initiative dialogue, iteratively refining recommendations by stating requirements such as ``the laptop should support video editing,'' ``smaller screen,'' or ``must have an integrated GPU.''

With each conversational turn that expresses new or refined requirements, the system re-ranks the product catalog and presents an updated product carousel within the chat history—always allowing users to revisit previous results.

% Seperate figures
\begin{figure}
  \centering
  \includegraphics[width=\columnwidth]{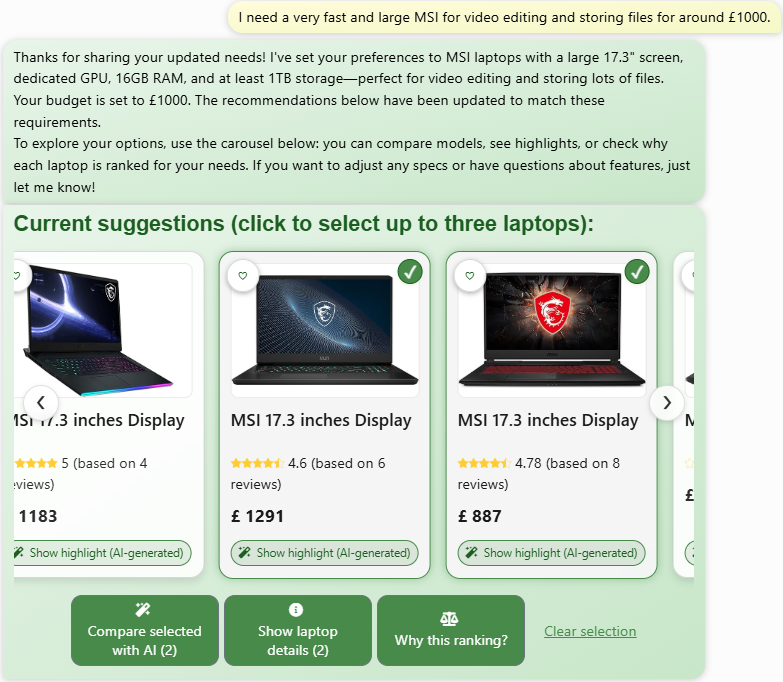}
  \caption{Product suggestions: If a user expresses new or refined needs, Cleo renders a suggestions carousel presenting the nine top-ranked items.}
  \label{fig:carousel}
  \Description{}
\end{figure}

\begin{figure}
  \centering
  \includegraphics[width=\columnwidth]{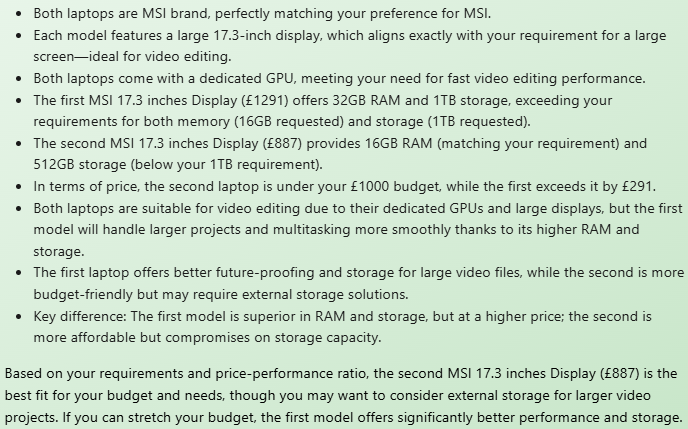}
  \caption{Multi-item analysis: selecting 2–3 cards enables the ``Compare selected with AI'' feature, which returns a bullet-pointed comparison plus a brief personalized recommendation.}
  \label{fig:comparison}
  \Description{}
\end{figure}

\begin{figure}
  \centering
  \includegraphics[width=\columnwidth]{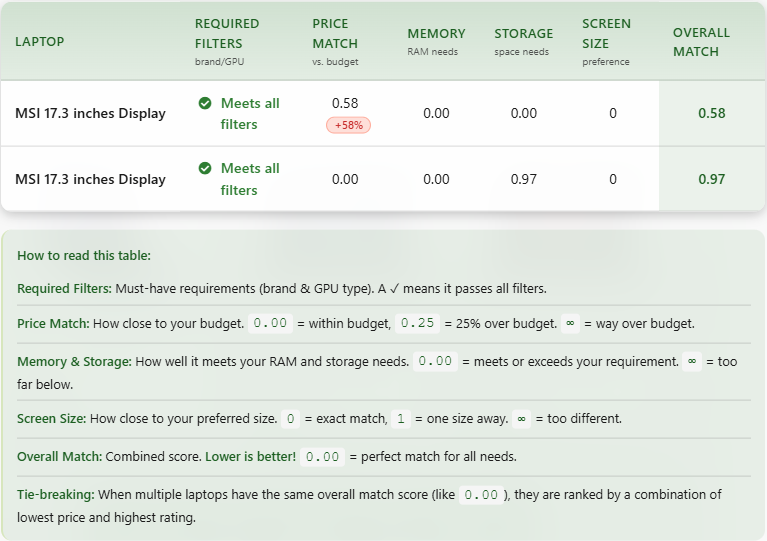}
  \caption{Explanations: The ``Why this ranking?'' modal displays required filters (brand/GPU), per-attribute loss values (price/RAM/storage/screen size), and an overall score (lower is better). It also shows a legend explaining the loss values.}
  \label{fig:ranking}
  \Description{}
\end{figure}

\subsection{Product Suggestions \& Highlights}
The nine top-ranked laptops are presented in a carousel view (see Figure~\ref{fig:carousel}), displaying a product image, star rating, and price. Each product card supports three actions: (i) selecting the item to view product specifications or enable AI-generated comparison, (ii) adding the item to a wishlist (heart icon) for later review, and (iii) triggering an AI-generated product highlight that explains how and why the product matches the user's information need, for example: ``This Lenovo laptop matches your preferred 14-inch portable size, offers more RAM (20 GB) and storage (1 TB) than you requested, and includes a dedicated GPU for enhanced performance.''

\subsection{Multi-item Comparison \& Details}
Users can select up to three laptops for comparison. Once at least two items are selected, the ``Compare selected with AI'' button generates a bullet-pointed comparison that takes the user's initial information need into account, followed by a concise personalized recommendation (see Figure~\ref{fig:comparison}). A complementary ``Show laptop details'' button opens a modal window containing a side-by-side product specifications table. The wishlist view replicates these features for saved items.

\subsection{Explanations: ``Why this ranking?''}
The ``Why this ranking?'' button opens a modal window explaining the system's ranking rationale (see Figure~\ref{fig:ranking}). The modal displays per-item required filters (brand and GPU type, indicated by ``\checkmark'' or ``\ding{55}'' symbols with human-readable reasons such as ``Meets all filters'') and numeric loss values for price, RAM, storage, and screen size, culminating in an overall score where lower values indicate better matches. The modal includes a legend explaining how to interpret each column.

\section{System Overview}
Cleo is implemented with a hybrid architecture that separates decision policy from language generation, addressing opacity and unpredictability in conversational search and recommendation. The system combines a deterministic, auditable ranking core with constrained LLM generation, enabling both conversational fluidity and algorithmic transparency.

\begin{figure*}[h]
    \centering
    \includegraphics[width = 1\textwidth]{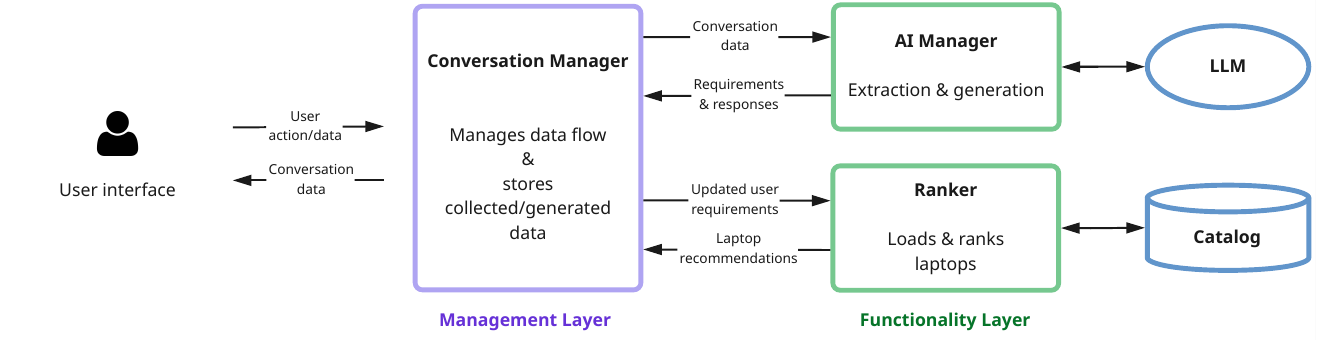}
    \caption{System architecture implementing our hybrid approach: the \textit{Functionality Layer} separates deterministic ranking (Ranker) from constrained language generation (AI Manager), while the \textit{Management Layer} (Conversation Manager) orchestrates their interaction.}
    \label{fig:Architecture}
    \Description{Architecture diagram showing three layers: server layer with REST API endpoints, management layer with conversation orchestration, and functionality layer with AI Manager and Ranker components}
\end{figure*}

\subsection{Architecture}
Figure~\ref{fig:Architecture} illustrates the backend architecture enforcing separation of tasks through two layers. The backend provides a REST API supporting message handling, product highlights and comparisons, and ranking explanations. The \textit{management layer} orchestrates dialogue flow through \textbf{Conversation Manager}, which invokes \textbf{AI Manager} and \textbf{Ranker} in sequence, and maintains the conversation state. The \textit{functionality layer} comprises two independent modules: \textbf{AI Manager} and \textbf{Ranker}. The \textbf{AI Manager} module operates in two modes: \textit{structured extraction} (anticipated user requirements in JSON format) and \textit{natural language generation} (responses, highlights, comparisons). Meanwhile, the \textbf{Ranker} implements a deterministic ranking, ensuring AI-generated conversational responses remain decoupled from predictable ranking decisions.

\subsection{Data Flow}
A typical interaction follows a four-stage process coordinated by \textbf{Conversation Manager}: (1) requirement extraction with (\textbf{AI Manager}), (2) deterministic re-ranking based on updated requirements (\textbf{Ranker}), (3) response generation using updated requirements and newly ranked laptops (\textbf{AI Manager}), (4) return of response and top-ranked items to the frontend.

Product comparisons and highlights use a simplified flow: hard-coded rules compute specification comparisons against requirements and across items, then the LLM formats them into bullet points through \textbf{AI Manager}. Ranking explanations use identical loss computations as the \textbf{Ranker} to ensure model-intrinsic transparency. User-chatbot conversations can optionally be logged.

\subsection{User requirements extraction}
A prompt instructs an LLM how the user's input can be mapped to structured JSON output with few-shot examples, e.g., ``casual gaming'' is mapped to \texttt{\{"gpu": "dedicated", "ram": 32\}}. It is supplemented by explicit scope instructions limiting generation to the laptop domain, and update policies that preserve previous values unless explicitly changed by the user. Following extraction, a sanitizer function corrects potential mistakes in the LLM's output through (1) unit conversion (e.g., converting terabytes to gigabytes or averaging price ranges), (2) aligning to catalog constraints (e.g., storage values to discrete tiers [256, 512, 1000, 2000~GB]), and (3) filtering invalid updates (e.g., if the LLM extracted a brand that is not in the catalog). This ensures extracted requirements map directly to queryable product attributes.

\subsection{Deterministic Ranking}
The \textbf{Ranker} module operationalizes system transparency through deterministic and inspectable ranking. An exemplary catalog of 3,638 laptop specifications is indexed for efficient computation. The ranking algorithm proceeds in three steps: Categorical filtering first removes devices that violate user requirements for brand and GPU type from the result list. Numeric loss functions then compute deviations for price, RAM, storage, and screen size. Finally, laptops are ranked by ascending total loss, considering both higher rating and lower price in case of loss value ties. Loss components are exposed through the ranking explanation endpoint, making every ranking decision auditable.

\begin{comment}
The \textbf{Ranker} module operationalizes system revealment through predictable, inspectable ranking. An exemplary catalog of 3,638 laptop specifications is indexed for efficient computation. The ranking algorithm proceeds in three steps:

\textbf{Step 1 - Categorical filtering:} Devices violating must-have requirements (brand/GPU) are eliminated from the result list.

\textbf{Step 2 - Loss computation:} Numeric loss functions compute deviations for price, RAM, storage, and screen size.

\textbf{Step 3 - Ranking:} Laptops rank by ascending total loss, considering both higher rating and lower price in case of loss value ties.

Loss components are exposed through the ranking explanation endpoint, making every ranking decision auditable.
\end{comment}

\subsection{Response Generation}
Another prompt instructs the system to articulate how vague user needs are mapped to concrete attribute values (e.g., ``For video editing, I've updated to a dedicated GPU''). Following RAG principles, responses are grounded in current user requirements and laptop recommendations to mitigate hallucinations. Few-shot examples demonstrate appropriate conversational style and handling of different user message types, e.g., for feedback, questions, or clarifications.

\subsection{Highlights and Comparisons}
For single-item product attributes, deterministic rules compare specifications to requirements (e.g., ``32 GB RAM exceeds 16 GB requirement''), while the LLM converts these into natural language descriptions. For multi-item comparisons, hard-coded rules first compare specifications across selected laptops and against requirements, then the LLM formats these as structured bullet points and adds a brief personalized recommendation based on price/performance trade-offs. This division aims to mitigate hallucinated specifications and over-persuasive content while allowing natural language comparison and contextual advice.

\section{Conclusion}
We demonstrated a transparent and controllable conversational product advisor addressing comparison complexity, opacity, and LLM unpredictability through a hybrid architecture. Our system provides transparency through auditable rankings and requirement reflection, controllability by separating deterministic ranking from language generation, and decision support through grounded natural-language comparisons and highlights. We contribute an extensible experimental system targeting IR/HCI researchers and practitioners seeking inspiration regarding the design of conversational search and recommendation systems.

Our approach differs from traditional faceted search (which lacks conversational interaction), end-to-end LLM recommenders (which typically hide decision logic), and existing comparison tools (which tend to present uncontextualized specifications). During the live demonstration, CHIIR attendees will be able to test their own search scenarios and provide feedback.

This work provides an adaptable pattern for building transparent and controllable conversational search systems in different domains. Current limitations include rigid loss weights and potential mental workload from the ranking explanations. We propose three potential user studies with Cleo: (1) comparing hybrid, LLM-only, and traditional faceted interfaces; (2) evaluating the explanation and decision support features to measure impact on decision confidence, trust, and time-to-decision; (3) exploring user preferences for guidance versus autonomy within a conversational search/recommendation system. Future work includes personalizing loss weights, expanding the product attributes taken into account, integrating customer reviews into item comparisons, and adding counterfactual steering controls.

We invite the CHIIR community to adapt this experimental system to advance transparency, control, and decision support in conversational search and recommendation.\footnote{The source code will be made publicly available upon publication at \url{https://osf.io/scxpv/overview?view_only=3dbccde002a846a5a13ab2001cd507b9}.}

\begin{acks}
This work was supported by the German Research Foundation (DFG) as part of the ``VACOS 2'' project (no. 388815326).
\end{acks}

%% BIBLIOGRAPHY
\bibliographystyle{ACM-Reference-Format}
\bibliography{main}

\end{document}